\documentclass[aps,prl,twocolumn,floatfix,showpacs,noshowkeys]{revtex4}%
\usepackage{amsmath,amssymb,bm,color,graphicx}

\newcommand{\ket}[1]{|{#1} \rangle}
\newcommand{\bra}[1]{\langle {#1}|}

\begin{document}
\preprint{MPI-PKS}
\title{Quantum phase transitions, entanglement, and geometric phases 
        of two qubits}
\author{Sangchul Oh}~\email{scoh@pks.mpg.de}
\affiliation{Max Planck Institute for the Physics of Complex Systems,
             N\"othnitzer Str. 38, D-01187, Dresden,
             Germany}
\date{\today}
\begin{abstract}
The relation between quantum phase transitions, entanglement, and geometric 
phases is investigated with a system of two qubits with XY type interaction.
A seam of level crossings of the system is a circle in parameter space of 
the anisotropic coupling and the transverse magnetic field, which is identical 
to the disorder line of an one-dimensional XY model. The entanglement of 
the ground state changes abruptly as the parameters vary across the circle except 
specific points crossing to the straight line of the zero magnetic field.
At those points the entanglement does not change but the ground state changes 
suddenly. This is an counter example that the entanglement 
is not alway a good indicator to quantum phase transitions. The rotation of the 
circle about an axis of the parameter space produces the magnetic monopole
sphere like a conducting sphere of electrical charges. The ground state evolving 
adiabatically outside the sphere acquires a geometric phase, whereas the ground 
state traveling inside the sphere gets no geometric phase.
The system also has the Renner-Teller intersection which gives no geometric
phases. 
\end{abstract}
\maketitle

Energy is the most primary quantity determining the properties of 
physical systems. When energy levels are crossing or avoided crossing
as system parameters vary, a quantum system exhibits rich physics.
For example, if two energy levels, initially separated, become close 
but not crossing and then far away again, then the non-adiabatic Landau-Zener 
transition between them takes place~\cite{Zener32}. 
Closely related to this, the runtime of adiabatic quantum computation is 
inversely proportional to the square of the minimum energy gap between 
the ground and first exited states~\cite{Farhi01}.
A quantum state traveling adiabatically around level crossing points 
accumulates a geometric phase in addition to a dynamical
phase~\cite{Berry84,Shapere89}. Berry put a beautiful interpretation on 
the geometric phase as the magnetic flux due to magnetic monopoles located 
at degenerate points. A quantum phase transition, a dramatic change in 
the ground state driven by parameters in zero temperature, is associated 
with a level crossing or avoided crossing between the ground and exited 
energy levels~\cite{Sachdev99,Vojta03}. 

Recently, entanglement, geometric phases, and fidelity have been adopted 
as new tools to characterize quantum phase transitions. Entanglement, 
referring to quantum correlations between subsystems, could be a good 
indicator to quantum phase transitions, because the correlation length 
becomes diverge at quantum critical points~\cite{Osterloh02,Osborne02,Amico08}.
Since the geometric phase and the quantum phase transition are associated 
with level crossing or avoided crossing, the geometric phase might be also 
used to characterizing quantum phase transitions~\cite{Carollo05,Zhu06}.
The fidelity, a measure of distance between quantum states, could be 
a good tool to study the drastic change in the ground states in quantum phase
transitions~\cite{Zanardi07}. So, it is natural to ask a question whether 
the entanglement, the fidelity, and geometric phases are always good indicators 
to quantum phase transitions or not.  If not so, why and when do 
they fail to characterize quantum phase transitions?  

In this paper, we give a partial answer to this question. A system 
of two qubits with XY type interaction is considered as a minimal model 
showing the quantum phase transitions, the abrupt change in the fidelity, 
the entanglement jump, and geometric phases. We present counter examples that 
the entanglement and geometric phases may fail to capture a certain quantum 
phase transition. In addition to this, it is shown that the magnetic 
monopole charges producing the geometric phase are distributed on the surface 
of the sphere in parameter space like a conducting sphere of electric charges.

An one-dimensional XY model with a large number of $1/2$ spins in 
a transverse magnetic field (hereafter called simply the XY model) is 
exactly solvable~\cite{Lieb61}, so it is a paradigmatic example in 
the study of quantum phase transitions.  Here we consider a simple system 
of two qubits with XY type interaction. As shown later, it contains rich 
physics in spite of its simplicity. The Hamiltonian of the system reads
\begin{align}
H(\lambda,\gamma)= 
&- \frac{(1+\gamma)}{2}\, \sigma^{x}_{1}\,\sigma^{x}_{2}
 - \frac{(1-\gamma)}{2}\, \sigma^{y}_{1}\,\sigma^{y}_{2} \nonumber\\ 
&- \frac{\lambda}{2}\left(\sigma^{z}_{1} +\sigma^{z}_{2} \right) \,,
\label{Hamil_XY}
\end{align}
where $\gamma$ is an anisotropy factor, $\lambda$ is an external magnetic field 
in the $z$ direction, and $\sigma_{i}^{\alpha}$ are the Pauli matrices of 
the $i$-th qubit with $\alpha = x,y,z$. 

The eigenvalues and eigenstates of Hamiltonian (\ref{Hamil_XY}) can be easily 
obtained by rewriting it in the matrix form,
\begin{align}
H(\lambda,\gamma) &= 
-\left(\begin{array}{cccr}
  \lambda & 0 & 0 & \gamma \\
  0       & 0 & 1 & 0\\
  0       & 1 & 0 & 0\\
  \gamma  & 0 & 0 &-\lambda
  \end{array}\right) 
= H_{\rm even} + H_{\rm odd} \,.
\label{Hamil_matrix}
\end{align}
The Hamiltonian 
$H_{\rm even} 
 = -\begin{pmatrix}
    \lambda & \gamma \\ 
    \gamma & -\lambda 
    \end{pmatrix}
$
is defined on the subspace spanned by $\{\ket{00},\ket{11}\}$. This looks 
like a Hamiltonian of a spin $1/2$ in a magnetic field. It is easy
to write down the eigenvalues and eigenvectors 
of $H_{\rm even}$
\begin{subequations}
\begin{align}
E_{\pm}^{e} &= \pm\sqrt{ \lambda^2 + \gamma^2}\,,
\label{ground_energy_even} \\
\ket{E_{-}^{e}} 
&= \cos\frac{\theta}{2}\, \ket{00} + \sin\frac{\theta}{2}\, \ket{11} \,, 
\label{ground_even} \\
\ket{E_{+}^{e}} 
&= -\sin\frac{\theta}{2}\, \ket{00} + \cos\frac{\theta}{2}\, \ket{11} \,,
\end{align}
\end{subequations}
where $\tan\theta \equiv \gamma/\lambda$. On the other hand, 
the Hamiltonian 
$H_{\rm odd} 
= - \begin{pmatrix}
    0 & 1 \\
    1 & 0
    \end{pmatrix}
$ acting on the subspace of $\{\ket{01},\ket{10}\}$ has the
eigenvalues $E_{\pm}^{o} = \pm 1$ and the eigenvectors 
\begin{align}
\ket{E_{\pm}^{o}} = \frac{1}{\sqrt{2}}\left(\ket{01} \mp \ket{10}\right) \,.
\label{ground_odd}
\end{align}

\begin{figure}[htbp]
\includegraphics[scale =1.0]{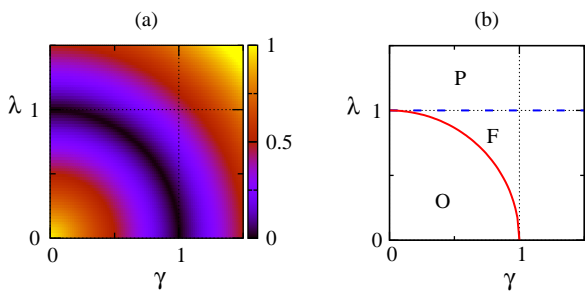}
\caption{(color online). (a) Energy gap between the ground and first
   excited states as a function of $\lambda$ and $\gamma$ for two qubits 
   with XY type interaction. The level crossing (dark line) takes places on 
   the circle $\lambda^2 + \gamma^2 =1$. (b) The ground state phase diagram 
   of the XY model.  P denotes the paramagnetic phase, F the ordered 
   ferromagnetic phase, and O the oscillatory phase~\cite{Barouch71}.}
\label{Fig1}
\end{figure}

Let us look at where the level crossings between the ground and first 
exited states are located in the parameter space of $\gamma$ and $\lambda$.
As shown in Fig.~\ref{Fig1}, the condition of level crossings, 
$E_{-}^{\rm o} = E_{-}^{\rm e}$, is just a circle, 
\begin{align}
\lambda^2 + \gamma^2 =1 \,. 
\end{align}
Surprisingly, this is identical to the disorder line of the  XY model,
which separates the ordered oscillating phase in the region 
$\lambda^2 + \gamma^2 <1$ from the ferromagnetic phase~\cite{Barouch71}.

It is remarkable that the ground state and the ground-state energy of 
the XY model are similar in forms to Eqs.~(\ref{ground_energy_even}) 
and (\ref{ground_even}), respectively. One may wonder why the phase diagram 
of two qubits with XY type interaction looks like that of the XY model 
in thermodynamic limit, as depicted in Fig.~\ref{Fig1}. This could be 
explained by the fact that the ground state energy $E_0$ of a system of $N$ 
identical particles with at most two particle interaction can be written
as $E_0= \frac{N}{2}\sum_{i} \epsilon_i \bra{\epsilon_i} D^2\ket{\epsilon_i}$
~\cite{Coleman}, where $D^2 = {\rm tr}_{3,\dots,N}\left( \ket{\Psi}\bra{\Psi}
\right)$ is the two-particle reduced density matrix derived from the ground
state $\ket{\Psi}$ satisfying $H\ket{\Psi} = E_0\ket{\Psi}$. 
And the reduced Hamiltonian $K\equiv H_1 + H_2 + (N-1)H_{12}$, derived from
the full Hamiltonian $H = \sum_{i=1}^{N} H_i + \sum_{i<j} H_{ij}$, has 
eigenvalues $\epsilon_i$ and eigenstates $\ket{\epsilon_i}$.
The XY model can be mapped to an one-dimensional spinless fermion system
through the Jordan-Wigner transformation. Thus one can see why
the eigenvalues and eigenstates of Hamiltonian~(\ref{Hamil_XY}) 
contain the partial information on the XY model~\cite{Oh08}. 

\begin{figure}[htbp]
\includegraphics[scale = 1.0]{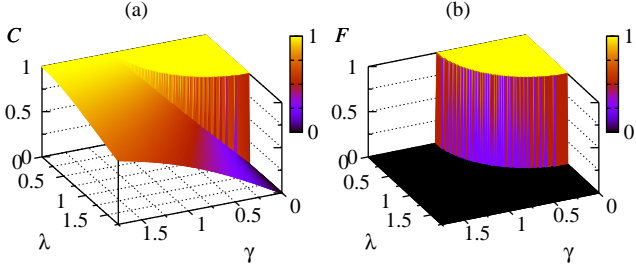}
\caption{(color online). (a) Concurrence $C(\ket{\psi_0})$ of the ground 
    state $\ket{\psi_0}$ and (b) fidelity $F$ between $\ket{E_{-}^{0}}$ and 
    $\ket{\psi_0}$ as a function of $\gamma$ and $\lambda$. }
\label{Fig2}
\end{figure}

Now, let us examine whether the entanglement is always a good indicator 
to quantum phase transitions~\cite{Osterloh02,Osborne02,Amico08}. 
For a pure two-qubit state $\ket{\psi} = a\ket{00} +
b\ket{01} + c\ket{10} + d\ket{11}$ with 
$|a|^2 + |b|^2 + |c|^2 + |d|^2 = 1$,
a well-known entanglement measure is the concurrence 
$C(\ket{\psi}) = 2|ad -bc|$. For the ground state $\ket{\psi_0}$
given by $\ket{E_{-}^{e}}$ and $\ket{E_{-}^{o}}$, it is written as
\begin{align}
C(\ket{\psi_0}) 
= \begin{cases}
  \sin \theta &\text{for}\quad \gamma^2 + \lambda^2 > 1\,, \\
  1           &\text{for}\quad \gamma^2 + \lambda^2 < 1\,. 
  \end{cases}
\end{align}
As shown in Fig.~\ref{Fig2} (a), the entanglement changes abruptly as 
$\gamma$ and $\lambda$ passes across the circle, i.e., the disorder line. 
It seems that the entanglement works well as an indicator to quantum phase
transitions. However, along the $\gamma$ axis, i.e., $\theta =\pi/2$, 
the concurrence doesn't change even if the ground state changes from 
$\ket{E_{-}^{o}}$ to $\ket{E_{-}^{e}}$. This demonstrates that 
the entanglement may fail to capture a certain quantum phase transition 
which happens between the ground states with same degrees of entanglement. 
As the number of particles increase, the dimension of a sub Hilbert space 
whose states have the same degree of entanglement also increases. It 
is not hard to imagine a quantum phase transition which occurs between 
the ground states with the same degree of entanglement. In this case, 
the entanglement is not a good indicator to quantum phase transitions. 

The fidelity $F$ between two quantum states $\ket{\psi} $ and $\ket{\phi}$,
defined by $F\equiv|\langle\psi |\phi\rangle|^2$, is one of the useful 
measures of distance between two quantum states. It could be a good indicator
to quantum phase transitions because the ground states before and after 
the quantum critical points change abruptly~\cite{Zanardi07}. 
It is simple to calculate the fidelity 
$F = | \langle E_{-}^{o} | \psi_0\rangle |^2$ 
between the ground state $\ket{\psi_0(\lambda,\gamma}$ and the reference 
state $\ket{E_{-}^{0}}$ as a function of $\gamma$ and $\lambda$. 
As illustrated in Fig. ~\ref{Fig2} (b), the fidelity  jumps on the circle. 

Let us turn to the relation between geometric phases and quantum phase
transitions. The geometric phase has been introduced as an alternative indicator 
to quantum phase transitions in Refs.~\cite{Carollo05,Zhu06}
where the degeneracy points are located on the XX line, i.e., along the 
$\lambda$ axis, of the XY model. In contrast, 
the system here has the degeneracy points on the circle. 
Let us rotate the Hamiltonian about the $\lambda$ axis by angle
$\phi$, 
$\widetilde{H}(\lambda,\gamma,\phi) 
= U^{\dag}_z(\phi)\, H(\lambda,\gamma)\, U_z(\phi)$
with $U_z(\phi) \equiv 
\exp\left[-i\frac{\phi}{2}\left(\sigma_{1}^{z} + \sigma_2^{z}\right)\right]$.
It is easy to obtain the transformed Hamiltonian  
\begin{align}
\widetilde{H}(\lambda,\gamma,\phi) = 
-\left(\begin{array}{cccc}
 \lambda & 0 & 0 & \gamma\,e^{-i2\phi} \\
 0       & 0 & 1 & 0 \\
 0       & 1 & 0 & 0 \\
 \gamma\,e^{i2\phi} & 0 & 0 & -\lambda
 \end{array}\right) \,.
\label{Hamil_trans}
\end{align}
The comparison of Eqs. (\ref{Hamil_matrix}) and (\ref{Hamil_trans}) 
shows two things. First, $\widetilde{H}_{\rm odd}$ is independent of 
angle $\phi$.  Second, $\widetilde{H}_{\rm even}$ looks like that of 
a spin $1/2$ particle in a rotated magnetic field 
${\bf B} = \sqrt{\lambda^2+\gamma^2}\, 
(\sin\theta\cos 2\phi,\sin\theta\sin 2\phi,\cos\theta)$. Notice that 
the azimuth angle is $2\phi$ even if the system is rotated by $\phi$.
This is due to the bilinear form of Hamiltonian~(\ref{Hamil_XY}). 
It is instructive to compare it with the rotation of a single spin about 
$z$-axis by $\phi$, as in a textbook of quantum 
mechanics~\cite{Sakurai}. The transformed Hamiltonian $\widetilde{H}_S$ 
is given by
\begin{align*}
\widetilde{H}_S 
=R^\dag_z \begin{pmatrix}  
    \cos\theta &\sin\theta\\
    \sin\theta &-\cos\theta
    \end{pmatrix}
    R_z
= \begin{pmatrix}
  \cos\theta  &\sin\theta\, e^{-i\phi}\\
  \sin\theta\,e^{i\phi} &-\cos\theta
  \end{pmatrix}\,, 
\end{align*}
where $R_z = \exp\left[ -i\frac{\phi}{2}\sigma_{z}\right]$.
This is $2\pi$ periodic in $\theta$ and $\phi$.
On the other hand, Hamiltonian (\ref{Hamil_trans}) is $\pi$ periodic in $\phi$, 
$\widetilde{H}(\theta,\phi) = \widetilde{H}(\theta+2\pi,\phi+\pi)$.

\begin{figure}[htbp]
\includegraphics[scale = 0.9]{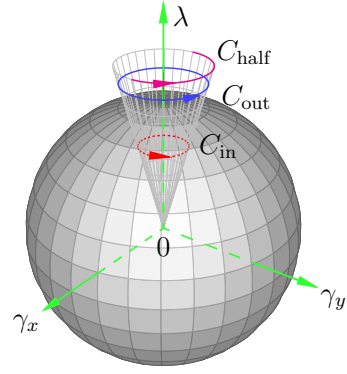}
\caption{(color online). A monopole sphere is produced by 
rotating a circle $\gamma^2 + \lambda^2=1$ about $\lambda$ axis 
by $2\pi$, where the monopole charges corresponding to the degeneracy points 
are distributed on the surface of the sphere, not located at the origin.
Two circuits, $C_{\rm in}$ inside and $C_{\rm out}$ outside the sphere, 
have the same solid angle. An open path $C_{\rm half}$ outside 
the sphere is a half rotation, i.e., $\pi$. 
In spherical coordinates, the radius $r$ is given by $r=\sqrt{\gamma^2
+ \lambda^2}$, the polar angle $\theta$ by $\tan\theta=\gamma/\lambda$,
and $\gamma = \sqrt{\gamma_x^2 + \gamma_y^2}$.
}
\label{Fig3}
\end{figure}

Let the ground state of the system be changed adiabatically 
through the path $C$ in the parameter space ${\bf r}=(r,\theta,\phi)$,
as shown in Fig.~\ref{Fig3}. The instantaneous ground state 
$\ket{\psi_0({\bf r})}$ satisfying $\widetilde{H}({\bf r})\ket{\psi_0({\bf r})} 
= E_0({\bf r}) \ket{\psi_0({\bf r})}$ reads
\begin{align}
\ket{\psi_0} 
= \begin{cases}
  \cos\tfrac{\theta}{2} e^{-i\phi}\, \ket{00} + \sin\tfrac{\theta}{2}
  e^{i\phi}\,\ket{11} 
  &\text{for}\,\, r > 1 \,,\\[6pt]
  \frac{1}{\sqrt{2}}\left(\ket{01} + \ket{10}\right)
  &\text{for}\,\, r < 1  \,.
  \end{cases}
\label{Ground_state}
\end{align}
For $r> 1$, the argument of the exponential function is $\phi$,  not $\phi/2$. 
With Eq.~(\ref{Ground_state}), it is easy to calculate the Berry connection 
or vector potential
${\bf A}_g \equiv i\bra{\psi_0({\bf r})} \nabla \ket{\psi_0({\bf r})}$,
\begin{align}
{\bf A}_{g} 
= \begin{cases}
  \displaystyle{
  -\frac{2}{r\sin\theta}\sin^2\frac{\theta}{2}\, \bm{\hat{\phi}}}
  &\text{for}\quad r > 1\,, \\[6pt]
  0           &\text{for}\quad r < 1\,. 
  \end{cases}
  \label{vector_potential}
\end{align}
In Eq.~(\ref{vector_potential}), the Berry connection outside the sphere is
two times stronger than that of a single $1/2$ spin in a magnetic field. 
It is evident that the ground state evolving adiabatically inside the sphere
acquires no geometric phase,
\begin{align}
\beta = \oint_{C_{\rm in}} {\bf A}_g\cdot d{\bf r} = 0 \,.
\end{align}
On the other hand, the ground state traveling outside the sphere accumulates 
the geometric phase 
\begin{align}
\beta = \oint_{C_{\rm out}} {\bf A}_g\cdot d{\bf r} =
-2\pi(1-\cos\theta)=-\Delta\Omega\,,
\end{align}
where $\Delta\Omega$ is the solid angle bounded by $C_{\rm out}$.
One may feel puzzled on why the geometric phase is twice 
of $\beta = -s\Delta \Omega$ of spin $s=1/2$ even if the quantum state 
outside the sphere looks like that of a single spin. The reason is that
the Hamiltonian $\widetilde{H}(\lambda,\gamma,\phi)$ has the period of 
$\pi$ in $\phi$. So, the quantum state traveling along the half-circuit 
$C_{\rm half}$ as shown in Fig.~\ref{Fig3} gets $\beta = -\pi(1-\cos\theta)$ 
but the path in the parameter space is not closed. This is closely related 
to the fact that a single spin returns to its original state after 
the $4\pi$ rotation, but the $2\pi$ rotation gives the minus
sign~\cite{Rauch75,Sakurai}.
In our case, consider the action of the operator $U_z(\phi)$
on the entangled state $\ket{\Phi^{(+)}}=\frac{1}{\sqrt{2}}(\ket{00} 
+ \ket{11})$. One obtains $U_z^\dag(\phi)\ket{\Phi^{(+)}}=\frac{1}{\sqrt{2}}
(e^{i\phi}\ket{00} + e^{-i\phi} \ket{11})$. Thus the entangled state
$\ket{\Phi^{(+)}}$ gets back to its original state after the $2\pi$ 
rotation but not $4\pi$.  It is possible to detect the global minus sign 
change in $\ket{\Phi^{(+)}}$ under the $\pi$ rotation.
Notice that a product state 
$\frac{1}{2}(\ket{0}_1 + \ket{1}_1)(\ket{0}_2 + \ket{1}_2)$ doesn't get
the global minus sign under the $\pi$ rotation of $U_z(\phi)$.
The magnetic monopole field  ${\bf B}_g = \nabla\times {\bf A}_g$
is written as
\begin{align}
{\bf B}_{g} 
=\left\{
  \begin{array}{cc}
  -\displaystyle{\frac{2}{r^2}\, {\bm\hat{\bf r}}} \quad
  &\text{for}\quad r > 1\,, \\[6pt]
  0\quad
  &\text{for}\quad r < 1  \,.
  \end{array}\right.
\end{align}
This is completely analogous to the electric field produced by 
a conducting sphere with charge $q=-2$. 

\begin{figure}[htbp]
\includegraphics[scale = 1.0]{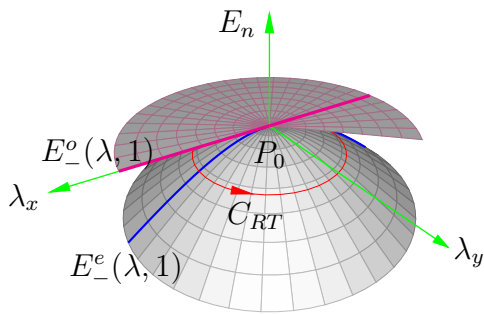}
\caption{(color online). Renner-Teller intersection at the point of 
   $\gamma=1$ and $\lambda=0$.
   }
\label{Fig4}
\end{figure}

Let us take a look at when the geometric phase fails to capture 
the quantum phase transition. Consider the level crossing along 
the line $\gamma = 1$ as shown in Fig.~\ref{Fig4}.
The ground state energy $E_{-}^{e}(\lambda,1) = -\sqrt{1+\lambda^2}$ meets 
the first exited energy $E_{-}^{o}(\lambda,1) = -1$ at $\lambda=0$,
but they don't cross each other. This is known as the Renner-Teller or 
glancing intersection, which differs from the conical intersection in 
the sense that the Renner-Teller intersection points is not a source of 
geometric phase~\cite{Zwanziger87}. To this end, let us rotate 
the Hamiltonian $ H(\lambda,1) = -\sigma_{1}^{x}\sigma_{2}^{x} 
 - \frac{\lambda}{2}\left(\sigma_{1}^{z} + \sigma_{2}^{z}\right)$
with $U_x(\phi) \equiv \exp\left[ -i\frac{\phi}{2}(\sigma^{x}_{1} 
+ \sigma^{x}_{2})\right]$. It is easy to show that two energy levels 
$E_{-}^{e}(\lambda,1)$ and $E_{-}^{o}(\lambda,1)$ have the cylindrical 
symmetry, i.e., independent of $\phi$, as depicted in Fig.~\ref{Fig4}.
The ground state of the transformed Hamiltonian
$\widetilde{H}(\lambda,\phi) = U_x^{\dag}(\phi) H(\lambda,1) U_x(\phi)$
is given by 
$\ket{\psi_0(\phi)} 
= (a\,\cos^2\frac{\phi}{2}-b\,\sin^2\frac{\phi}{2})\,\ket{00} 
+ i\frac{a+b}{2}\sin\phi\,\ket{01} 
+ i\frac{a+b}{2}\sin\phi\,\ket{10} 
+ (b\,\cos^2\frac{\phi}{2}-a\,\sin^2\frac{\phi}{2})\,\ket{11}$,
where $a =\cos\frac{\theta}{2}, b=\sin\frac{\theta}{2}$, and
$\tan\theta=1/\lambda$. Clearly there is no sign change 
in $\ket{\psi_0(\phi)}$ under the rotation of $\phi$ by $2\pi$.

Let us discuss an experimental aspect of the main results presented here. 
Since the system considered here is composed of only two qubits, it may 
not be hard to verify them with NMR qubits~\cite{Jones00,Zhang08}, 
ion-trap qubits~\cite{Leibfried03}, superconducting qubits~\cite{Leek07}. 
With these systems, the geometric phase of a single qubit has been already 
reported. 

In conclusion, we have studied the system of two qubits with XY type 
interaction to investigate whether the entanglement, the fidelity, and 
the geometric phases are good indicators to quantum phase transitions. 
First, it has been shown that the phase diagram of two qubits with 
XY type interaction is same to the disorder line in the XY model. 
Second, we have presented the counter examples that 
the entanglement and the geometric phase fail to detect a certain 
quantum phase transition. As shown before, there is a quantum phase 
transition between the ground states with same degrees of entanglement.  
A quantum state traveling around the quantum phase transition point 
of the Renner-Teller intersection acquires no geometric phase. 
Finally, we have found the magnetic monopole sphere, only outside 
which the quantum state acquires the geometric phase, 
in perfect analogy to a charged conducting sphere. 

\acknowledgments
This work was supported by the visitor program of Max Planck Institute for the
Physics of Complex Systems.

\end{document}